\documentstyle[aps,prb,preprint]{revtex}
\begin{document} 
\draft
\title{\bf Two Interacting Electrons in a Quasiperiodic Chain}
\author{{\bf S.N. Evangelou and D.E. Katsanos} \\
 {\em Department of Physics}\\
 {\em University of Ioannina}\\ 
 {\em Ioannina 451 10 Greece}}
\date { }
\maketitle
\begin{abstract} 
 We study numerically the effect of  on--site Hubbard
 interaction $U$ between two electrons in the quasiperiodic
 Harper's equation. 
 In the periodic chain limit by mapping the problem
 to that of one electron in two dimensions with a diagonal
 line of  impurities of  strength $U$ we demonstrate  a band of
 resonance two particle pairing states starting from $E=U$.
 In the ballistic (metallic) regime we show 
 explicitly  interaction--assisted extended pairing states
 and multifractal  pairing states in the  diffusive (critical) 
 regime.  We also obtain localized pairing states
 in the gaps and the created subband due to $U$, whose number
 increases when going to the localized regime, which are
 responsible for  reducing the 
 velocity and the diffusion coefficient in the 
 qualitatively similar to the non--interacting case
 ballistic and diffusive dynamics. 
 In the localized regime we find propagation enhancement
 for small $U$ and stronger localization for larger $U$,
 as in disordered systems.
\end{abstract} 
 
\newpage\clearpage

 \section {Introduction}

 Anderson localization \cite{1}  can be also studied in
 quasiperiodic systems via the Harper's equation, which
 also describes electrons in a square lattice with an added
 strong magnetic field \cite{2,3,4}, superconducting
 networks \cite{5}, etc.
 This model presents  a very  useful alternative to the study
 of one--dimensional ($1D$) disordered systems, since  apart from
 localization  it can also display  metallic behavior
 associated with ballistic motion and critical behavior
 (mobility edge) with ordinary  diffusion \cite{6,7}, somehow
 mimicing  more realistic three--dimensional ($3D$)
 disordered systems. In this paper we
 report results which relate to the problem of what happens
 to the electronic eigenstates and the corresponding quantum 
 dynamics of two electrons
 moving in a  quasiperiodic potential in which the
 interaction between them is  taken into account. 

 The study of two Hubbard interacting particles (THIP)
 localized by a  random potential has been pioneered by
 Shepelyansky \cite{8}. This author and also 
 \cite{9,10,11,12,13,14} produced very interesting
 analytical and numerical  work within the 
 Anderson-Hubbard model, which showed weakening
 of Anderson localization, which is known 
 always to be caused by disorder
 in $1D$, due to the  effect of  the two particle interaction.
 This lead  to an enhanced propagation effect 
 of the interacting 
 electron pair on scales  larger  than the  single--particle
 localization length. The phenomenon of propagation enhancement    
 due to the interaction was also displayed in disordered 
 mesoscopic rings threaded by magnetic  flux 
 by showing a pairing effect via a few  $h/2e$--periodic,
 instead of $h/e$--periodic, eigenstates \cite{11}.
 However, a previous diagonalization study for $1D$
 disordered system  \cite {15} 
 revealed that a few states in the main band show  a
 weak propagation enhancement while 
 states with two locally paired  electrons are, usually, 
 even more localized in the presence of the interaction. 
 Moreover, it was pointed out that for large positive $U$ 
 stronger localization occurs when compared to the 
 non--interacting case.  In  a recent work \cite{16},
 it is clearly demonstrated by a tranfer matrix study 
 that no propagation enhancement is possible  for 
 THIP in an infinite disordered chain at $E=0$. 

 These, rather conflicting results can be partially
 understood by the fact that most  works, apart from 
 the direct exact diagonalization or dynamical approaches, 
 rely on a mapping of the  THIP problem to a superimposed 
 banded random matrix ensemble (SBRME). It was 
 suggested that if the interaction is expressed
 in the non--interacting localized 
 basis  a random band matrix
 with additional  disorder in the matrix diagonal 
 appears (SBRME) and the enhancement of the pair
 localization length $\xi \propto \xi_1^{2}$,
 where $\xi_1$ is the one--particle 
 localization length,  is easily obtained \cite{8}. However, 
 the reduction to a SBRME relies on a
 questionable assumption  about chaoticity of the 
 non--interacting localized states within $\xi_1$, so that the
 relevant matrix model could be probably different \cite{17}. 
 Moreover, since in the  one--particle localized basis
 the interaction is relevant only when the two particles
 are localized  around positions close to each other, 
 the obtained localization weakening 
 might vanish for an infinite chain \cite{16}. 
 It must be also pointed out that since in most of the
 previous works only the localized case with finite $\xi_1$
 is considered it could be reasonable to expect more 
 dramatic pairing effects for extended non--interacting
 eigenstates which are always overlapping. The
 model considered in this paper 
 allows to study 
 the fate of extended and critical one--electron states in the 
 presence of Hubbard interaction. We find a kind of 
 pairing effect  for the two particle states  in the
 metallic and the critical regimes with 
 the simultaneous appearance
 of localized pairing states in the gaps and the created 
 subband  due to the interaction \cite{18}.
 These  localized states
 reduce the corresponding THIP dynamics although it remains
 similar in nature to the non--interacting case in all 
 three regimes. However, we obtain a weak enhancement  
 of propagation in the localized regime when the interaction 
 is switched on, as in disordered systems, 
 but even  stronger localization is shown to occur
 for larger $U$.

 In sections II, III we  introduce the Harper-Hubbard model 
 and consider by diagonalization methods two interacting
 electrons moving in a tight binding  quasiperiodic  potential 
 of strength $\lambda$, for various values  of the local
 electron--electron Hubbard interaction $U$. 
 In the absence of the interaction $U$ it is known \cite{4} that
 there are extended states
 for $\lambda <2$, a mobility edge 
 for $\lambda = \lambda_{c} =2$ and a finite one--electron
 localization length $\xi_1=1/\log (\lambda /2)$ 
 independent of energy  for $\lambda >2$.
 In section IV  we present our results from the 
 numerical diagonalization of  the corresponding 
 two interacting electrons Harper--Hubbard 
 equation, by showing  explicitly  extended pairing states
 in the metallic case  $\lambda <2$ 
 or multifractal pairing states at the mobility edge $\lambda =2$.
 We do not find extended or multifractal pairing states 
 for the insulator but  a weak
 propagation enhancement with the 
 simultaneous appearance of  localized
 pairing states \cite{15}. In section V we
 address the question of the electron  localization
 dynamics in the presence of $U$. The  time evolution of 
 a quantum wave packet in the presence of interactions 
 shows  ballistic motion for $\lambda<2$, diffusion for
 $\lambda=2$ and ceases to expand for $\lambda >2$
 as for the non--interacting metallic,
 critical and localized regimes, respectively \cite{6,12}.
 In the presence of the interaction 
 a decreasing $U$--dependent electron velocity and 
 diffusion coefficient due to the appearance of 
 localized states is obtained for the metallic and the 
 critical regimes, respectively. In the localized
 regime  larger localization length is found for small $U$
 although for higher $U$ stronger localization occurs,
 in agreement with previous results on the Anderson-Hubbard model
 \cite{15}. Finally, in section VI we discuss our results 
 and present the conclusions which arise from the present study.

 \section {The Harper-Hubbard Model}

 The Harper-Hubbard tight binding 
 equation for two interacting particles is \cite{19,20} 
 \begin{eqnarray}
 H & = & \sum_{n=1}\sum_{\sigma }(c^{\dag }_{n+1,\sigma }c_{n,\sigma } 
 +c^{\dag }_{n,\sigma }c_{n+1,\sigma } )
 +\sum_{n=1}\sum_{\sigma }
 \lambda \, \cos(2\pi\phi n) c^{\dag }_{n,\sigma }
 c_{n,\sigma }  \nonumber \\
 &      & \mbox{}
 +\sum_{n=1}
 U c^{\dag }_{n,\uparrow}c_{n,\uparrow }
  c^{\dag }_{n,\downarrow }c_{n, \downarrow },
 \end{eqnarray}
 where $c^{\dag }_{n,\sigma }$ and $c_{n,\sigma }$ are the creation and 
 destruction operators for the electron at site $n$ with spin 
 $\sigma $, $\lambda \cos(2\pi\phi n)$ is the potential
 at site $n$, with $\phi$  an irrational number usually chosen
 as the golden mean $\phi = {\frac {\sqrt5-1}{2}}$  \cite{4} and  
 $U$ is the strength of the local Hubbard interaction 
 between the two electrons. 
 The Hilbert space can be conveniently divided into one 
 singlet subspace with total spin $S=0$ and three triplet
 subspaces  with total spin $S=1$, $S_z=1,0,-1$, respectively.
 The three triplet subspaces are energy degenerate and since 
 they permit no double occupation the triplet states 
 are not affected by the Hubbard
 interaction.  In a chain of $N$ sites the singlet subspace is
 spaned  in the basis of $N(N+1)/2$  spatially symmetric wave
 functions
 \begin{equation}
 \label{bas}
 |\psi (n_1,n_2)>_s = \left\{ 
 \begin{array}{l} 
 \frac{1}{\sqrt{2}}(c^{\dag }_{n_1,\uparrow }c^{\dag }_{n_2,
 \downarrow }
 +c^{\dag }_{n_2,\uparrow }c^{\dag }_{n_1,\downarrow })|0>
 \ \mbox{for }
 n_1\ne n_2, \\ 
 c^{\dag }_{n_1,\uparrow }c^{\dag }_{n_1,\downarrow }|0> \ 
 \mbox{for } n_1=n_2,
 \end{array} \right.
 \end{equation}
 which are antisymmetric 
 with respect to the exchange of the spins and permit 
 double occupancy.

 \section {Method of Calculation}

 We carried out exact diagonalization of $H$ in the 
 singlet subspace where the Hubbard interaction is
 relevant and found out all the eigenvalues and eigenvectors 
 for finite $N$ sites with various $\lambda$'s and $U$'s.
 In order to measure the degree of localization for the
 interacting electrons we calculate the one--particle spatial
 extent $\xi^{(j)} $ in the $j$th two--electron wave function via
 \cite{15}
 \begin{equation}
 \xi ^{(j)}=\sum_{n_1=1}^{N}\sum_{n_2=1}^{n_1}
 |a_{n_1,n_2}^{(j)}|^2
 \sqrt{(n_1-\overline{x}_1)^2+(n_2-\overline{x}_2)^2},
 \end{equation}
  with mean positions 
 \begin{equation}
 \overline{x}_{1,2} =\sum_{n_1=1}^N\sum_{n_2=1}^{n_{1}}
 |a_{n_1,n_2}^{(j)}|^2 n_{1,2}
 \end{equation}
 of the electrons $1$ and $2$, where
 $a_{n_1,n_2}^{(j)}$ is the normalized coefficient of the wave
 function  in the basis of Eq. (\ref{bas}). 
 It must be mentioned that in the way $\xi $ is defined
 it can estimate the  spatial extend of each electron 
 averaged over the second electron and is
 related to a quantity known as the participation ratio \cite{15}.
 Moreover,  $\xi $ should 
 correspond  to the true localization length if the 
 wave functions decay exponentially.
 Another important quantity used in this study is 
 the mean value of the
 distance  between the two electrons in the chain which can be
 calculated for each two--particle wave function via 
 \begin{equation}
 d^{(j)}=\sum_{n_1=1}^N\sum_{n_2=1}^{n_1} |a_{n_1,n_2}^{(j)}|^2
 |n_1-n_2|.
 \end{equation}
 The distance  $d $ measures the correlation between the two 
 electrons  so that a small $d$ 
 defines a pairing  two--electron eigenstate, which can be 
 either delocalized in the metallic regime, multifractal
 in the critical regime or localized mostly in the insulating 
 regime $\lambda >2$.

 \section {Two Particle Pairing States}

 We diagonalize the Hamiltonian matrix for 
 Fibonacci number chain lengths $N$, e.g. $N=89$ if 
 the rational approximant of $\phi = 
 {\frac {\sqrt5-1}{2}}$ is ${\frac {55} {89}}$,
 so that the potential is periodic with period $N$.
 In Fig. 1(a) we plot  $\xi$ and $d$ 
 versus the corresponding eigenvalue $E$
 for the $1D$ pure $\lambda =0$  case with interaction $U=1$. 
 The striking characteristic  is  a band of $N$ states,
 out of the total of $2N^{2}$ two--particle states,
 which have extremely  small  distance $d$
 starting from the energy $E=U$ where $d$ is presisely zero
 (see Fig. 1(a)).
 These   pairing states have one sharp peak at the diagonal
 line of the plane where $n_{1}=n_{2}$ which implies that the 
 particles always stay very close to each other. In Fig.
 1(b), (c), (d), (e) we plot some
 characteristic such wave function amplitudes  
 in  the plane of the two--electron coordinates $n_1$ and $n_2$
 where the  non--zero amplitudes appear in 
 or very close to the diagonal line. 

 Extended pairing states due to the interaction but having
 a finite width $d$ are also seen in Fig. 2  for  the ballistic 
 case $\lambda = 1$ for $ U=1$. These states are identified 
 from Fig. 2(a) by plotting in Fig. 2(b), (c)
 only some states which have small
 $d$. The extended pairing
 states in the metallic  regime ($\lambda <2$) 
 have  their number progressively reduced 
 when increasing  $\lambda$ towards 
 $\lambda= 2$. Fig. 3(a) accounts for the critical case 
 $\lambda=2$ where still a few pairing states are
 seen, such as in Fig. 3(b), with a displayed
 kind of multifractality along the diagonal \cite{7}.
 However, apart from extended or 
 multifractal pairing states  we also obtain another kind
 of localized pairing
 states which occur in pairs of almost identical energies
 and similar amplitude distributions.  In Fig. 3(c)  one such
 state is shown where a double peaked structure is displayed 
 along the diagonal having 
 small $d$ and misleadingly large $\xi$  
 due to our definition of $\xi$, since 
 such pairing states are strongly localized 
 in two spatial positions along the diagonal. The
 localized pairing states  correspond  to a physical
 picture of localization
 due to Mott \cite{21} and they are  more 
 frequently encountered  in the insulating  $\lambda>2$ 
 regime. They involve
 tunneling transitions between the two particle localized states
 spaced at a distance proportional to 
 $\xi_1$ apart, having energies that differ by very small
 amounts. In Fig. 4(a) we demonstrate $\xi$ and $d$  
 in the critical regime $\lambda =2$ with
 a higher value of the interaction strength $U=5$.
 In Fig. 4(b), (c) we show two multifractal pairing states. In 
 the  plot of Fig. 5 we show $\xi$ and $d$ 
 for the insulating regime  $\lambda =3$ with 
 localized pairing states having 
 two maxima (Fig. 5(b), (c)) and no extended 
 or multifractal pairing states survive  in this case.

 We find that localized
 two--particle states due to the interaction
 also appear in  the metallic and the critical regimes. These
 pairing states have small $d$ and are located either in the  
 gaps or in the subband created by the interaction
 $U$. They are identified from Fig. 2(a), 3(a), 4(a), 5(a)
 and  for the critical $\lambda =2$ case also
 in  Fig. 6(a), (b), (c) from the integrated density of 
 states which  is known to  be  multifractal ``devil's 
 staircase"  for non--interacting electrons. 
 For the  THIP four major gaps (plateus)
 are seen to coexist  with smaller gaps on all scales.
 In the created subband for large positive energy 
 the localized pairing states due to the  effect of the
 interaction $U$ are clearly seen.  
 It must be emphasised that localized pairing states are
 found for $\lambda <2$ 
 only in the presence of finite interaction
 ($U>0$) and the results 
 described in this section did not change
 qualitatively  by varying the system size.

 \section {Two Particle Dynamics}

 The study of the wave packet dynamics provides a global 
 information for the changes due to the interaction of all
 the relevant wave functions. If we put two electrons 
 at the same initial site, e.g. the chain center $0$  at $t=0$,
 the mean square displacement $<(\Delta x(t))^2>$
 for each electron at subsequent times $t$ can be calculated 
 from all the singlet eigensolutions  of Eq. (1) 
 from the variance
 \begin{equation}
 <\Delta x^2(t)> = \frac {1} {2} <n_1^2+n_2^2>
 =\sum_{n_1=1}^N\sum_{n_2=1}^{n_1}
 \left| \sum_j e^{-iE_jt}a^{(j)*}_{0,0}a^{(j)}
 _{n_1,n_2}\right| ^2(n_1^2+n_2^2)/2,
 \end{equation}
 where $<...>$ denotes quantum average and the factor 
 of $2$ in the denominator transforms $<\Delta x^2(t)>$
 to correspond to one  electron, in order to agree  with 
 previous one--electron dynamics  for $U=0$. 
 For the adopted initial condition, in which  
 the two electrons are at the same site, only the singlet
 states with additional on--site energy due to $U$ are allowed.
 Alternatively, we have 
 integrated the corresponding two--dimensional
 equations of motion using a Runge-Kutta algorithm, in order
 to obtain results for much longer chains of $N=17711$. Fig. 7(a) 
 shows the obtained $<(\Delta x(t))^2>$ for $\lambda =1$
 where  is seen that the  ballistic motion $<(\Delta x(t))^2>
 \propto t^2$, remains valid also for finite $U$ 
 but  with a reduced velocity. In the critical case $\lambda =2$  
 diffusion with $<(\Delta x(t))^2> \propto t$ is obtained  in
 Fig. 7(b) reducing in magnitude by increasing $U$, although 
 a tendency for more enhanced propagation is seen when
 $U=5$.  For the insulator in Fig. 7(c)
 $<(\Delta x(t))^2>$ shows  many oscillations and asymptotically
 reaches larger values for finite $U=1, 5$
 when compared to $U=0$,  which 
 indicates the familiar weakening of localization due to the interaction
 \cite{9,10,11,12,13,14}.
 However, for very large $U=7, 10$ the relevant wave 
 functions for the dynamical process lie mostly in the 
 subband created above the main band,
  which corresponds to localized pairing 
 states having much shorter localization lengths, and
 as a result the  mean square displacement becomes very short
 (fig. 7(c)) indicating a localizing effect of the interaction
 \cite {15}. 
 Therefore, a decrease of the  
 degree of localization due to the interaction 
 is demonstrated in the localized regime for not too large $U$,
 in agreement with the original reported tendency. 

 It must be mentioned that our definition of Eq. (6) focuses on
 the properties of one interacting electron and is
 different from both $\sigma^{2}_+
 = \frac {1} {4} <(n_1+n_2)^2> $ 
 and $\sigma^{2}_-= <(n_1-n_2)^2> $ introduced \cite{8,18}
 to examine coherent propagation of two electrons. Our 
 results  diplayed in Fig. 8 for the metallic, critical and the 
 localized cases show  a similar behavior of the electron
 ($\sigma^2$) and  pair ($\sigma^{2}_+$) propagation as well as 
 for the squared pair size
 ($\sigma^{2}_-$). From these
 results since the obtained propagation behavior
 in the diagonal and its  vertical are similar no coherent pair
 propagation can be concluded although a kind of
 weak pairing can be  seen  for the metal and the insulator
 where  $\sigma^2_+ > \sigma^{2}_-$ but no such effect at
 the critical point where  $\sigma^2_+ \simeq \sigma^{2}_-$.
 Moreover, in the insulating regime for $U=1$ (Fig. 8(c))
 we note that $\sigma^2$ is below  $\sigma^{2}_+$.  

 \section {Discussion - Conclusions}

 It can be shown \cite{8,10} that the interacting
 electron problem in the  periodic $\lambda=0$ case can be 
 mapped onto an equation for a single electron   moving in a two 
 dimensional lattice with a line of impurities of energy $U$ along
 the diagonal. The impurities naturally lead to $N$
 resonance states  at energies starting from  $E=U$ 
 as seen in Fig. 1(a),
 having amplitude only  on the impurity sites along the lattice
 diagonal. In this way  extended pairing states naturally 
 appear, for example,  precisely at $E=U$ a two--particle 
 pairing state can be found exactly (Fig. 1(b)) having a constant
 amplitude on the diagonal and zero elsewhere. 
 It is very well known that such resonance extended 
 states can also appear at certain energies 
 in non--interacting $1D$ chains
 with distributed  large 
 segments  of identical impurities \cite{22}.

 The quasiperiodicity in addition 
 to the interaction also permits such an exact mapping to a
 single electron equation  moving in a two 
 dimensional lattice with  a symmetric potential 
 $\lambda \, \cos(2\pi\phi n_1) +
 \lambda \, \cos(2\pi\phi n_2)$ at the 
 coordinate  $n_1, n_2$, which denote the positions of the
 two electrons, in addition to the 
 line of impurities $U\, \delta_{n_1,n_2} $ along the diagonal.
 In analogy with the one--dimensional large impurity 
 case where  additional small perturbations are known to 
 allow the survival of a weak resonant 
 effect at certain energies, remnant of the extended states in the 
 absence of perturbation \cite{22}, we similarly
 obtain a kind of pairing states having 
 finite but small distance between the two electrons.
 In this paper we demonstrate  by 
 exact diagonalization of the THIP 
 Hamiltonian in a finite quasiperiodic
 Harper's chain such two--particle extended or multifractal
 pairing states due to the interaction
 for the metal and at the critical point.
 In these regimes  we also find localized pairing states
 due to the interaction in the gaps or the created
 subband, also according to the Mott resonance theory 
 of localization. In the localized regime $\lambda>2$
 we find mostly  localized pairing states with short
 localization lengths.

 In disordered systems previous attempts to consider the
 electron-electron interactions are based on perturbation
 theories \cite{23,24} or more rigorous solutions  for  special
 cases \cite{25,26}. In this paper by a numerical diagonalization
 study the interaction between two electrons 
 a novel pairing effect for certain delocalized states is found
 which occurs via extended or multifractal pairing
 states. Localized pairing states are also found due
 to the interaction in all regimes \cite{18}.
 Our results are confirmed 
 for both repulsive and attractive on-site Hubbard 
 interactions and is probably
 worth noting that in the case  
 of an attractive  interaction ($U$ negative)  the
 subband created by the  interaction lies below the band bottom
 of the non-interacting case so that the  ground state is
 always a localized pairing state for the metal and 
 the insulator. The localized pairing states might 
 also affect physical quantities since they can 
 decrease the  velocity and the diffusion coefficient in the
 metalic and the critical regimes. Moreover,
 in the corresponding dynamics we demonstrate
 a tendency for weakening of localization in the 
 insulating $\lambda >2$ regime for small $U$ 
 but even stronger localization for higher $U$.

 {\Large\bf Acknowledgments} 

 We would  like to thank J.-L. Pichard for originally 
 introducing us to the problem, S.J. Xiong, E.N. Economou and 
 C.J. Lambert for many  useful discussions.
 This work was supported in part by a $\Pi$ENE$\Delta$ Research 
 Grant of the Greek Secretariat of Science and Technology,
 from EU contract CHRX-CT93-0136 and within a TMR network.

 \newpage\clearpage

\begin{figure}
    
 {\bf Fig. 1.}   {\bf (a)} The electron spatial extend $\xi $ 
 (open circles) and the mean distance $d$ (black dots)
 between  two electrons as a
 function  of the two-electron wave function  energy
 $E $ for the periodic Hubbard chain of $\lambda=0$ 
 with  interaction strength $U=1$.
 {\bf (b),(c),(d),(e)} Amplitude  distributions for  
 extended pairing states with $E=U$, $E=4.1225$, $E=2.5119$, and
 $E=4.1173$, respectively, as a function of the two electron
 coordinates $n_1$ and $n_2$ 
 and the same parameters as in (a).
 \\
 {\bf Fig. 2.}  {\bf (a)} The $\xi $ 
 and $d $ versus the energy
 $E $ of the two-electron wave functions in
 the {\bf ballistic} case
 $\lambda=1$ with the interaction strength $U=1$.
 Amplitude  distributions {\bf (b)} for extended pairing 
 states $E=-0.0013$ and
 {\bf (c)}  for $E=-2.5707$.
 \\
 {\bf Fig. 3.}   {\bf (a)}  The $\xi   $ 
 and $d $ versus the energy
 $E $ of the two-electron wave function for the 
 {\bf diffusive}
 case with $\lambda=\lambda_c=2$ and
 the interaction strength $U=1$.
 Amplitude  distributions {\bf (b)} for pairing states
 $E=-0.5231$  has a  multifractal character and {\bf (c)}
 for $E=0.8274$ is localized pairing state in the Mott sense.
 \\
 {\bf Fig. 4.}   {\bf (a)}  The $\xi  $ 
 and $d $ versus the energy
 $E $ of the two-electron wave function for the critical
 {\bf diffusive} case $\lambda=2$ with $ U=5$. The states
 {\bf (b)}  $E=1.3631$ and {\bf (c)}  $E=1.3752$ display
  a  multifractal character.
 \\
 {\bf Fig. 5.}   {\bf (a)}  The $\xi  $ 
 and $d $ versus the energy
 $E $ of the two-electron wave function for the insulating
 {\bf localized} case $\lambda=3$ with $ U=1$. The 
 {\bf (b)}  $E=-0.0646$ and {\bf (c)}  $E=0.9083$ correspond
 to localized pairing states  in the Mott sense.
 \\
 {\bf Fig. 6.}  {\bf (a), (b), (c)}
 The integrated density of states for the 
 critical case $\lambda =2$ with various values of the 
 interaction strength $U$. The main gaps correspond to the 
 plateaus.
\newpage\clearpage
 {\bf Fig. 7.}  The mean square displacement $<(\Delta x(t))^2>$ 
 of a wave packet for the 
 approximant $\phi={\frac {10946}{17711}}$ 
 with the two electrons initially located at
 the chain center  of long length  $N=17711$  which ensures
 that the wave does not reach the ends of the chain.
 The values of the interaction strength $U$ are denoted
 in the figures:
 {\bf (a)}  log-log plot for the {\bf ballistic} case $\lambda =1$,
 {\bf (b)}  log-log plot for the {\bf diffusive} case  $\lambda =\lambda_c
  =2$ and {\bf (c)}  ordinary plot  for the {\bf localized} case
  $\lambda =3$.
 \\
 {\bf Fig. 8.} 
 A comparison between the mean square displacements
 $\sigma^2 = <(\Delta x(t))^2>$, the pair mean square 
 displacement $\sigma^2_{+}$ and the pair size 
 $\sigma^2_{-}$ with the rest of parameters as in Fig. 7.
 The values of the interaction strength $U=1$:
 {\bf (a)}  log-log plot for the {\bf ballistic} case $\lambda =1$,
 {\bf (b)}  log-log plot for the {\bf diffusive} case  $\lambda =\lambda_c
  =2$ and {\bf (c)}  ordinary plot  for the {\bf localized} case
  $\lambda =3$.
\end{figure} 
 \end{document}